\documentclass[journal=jacsat,manuscript=article]{achemso}

\usepackage[version=4]{mhchem} 
\usepackage[utf8]{inputenc}
\usepackage{textcomp}
\usepackage{graphicx}
\usepackage{siunitx}
\DeclareSIUnit\au{a.u.}
\usepackage{amsmath}
\usepackage{upgreek}
\usepackage{hyperref}
\usepackage{amssymb}
\usepackage{tabularx}
\usepackage{booktabs}
\usepackage{epstopdf}
\AppendGraphicsExtensions{.tiff}
\usepackage{comment}
\usepackage{xcolor}
\colorlet{review}{red}
\DeclareUnicodeCharacter{0301}{\'}
\allowdisplaybreaks

\author{Megan Grace}
\affiliation{Department of Chemistry, University of Dayton, OH, USA}
\author{Avdhoot Datar}
\affiliation{Department of Chemistry, University of Dayton, OH, USA}
\email{adatar1@udayton.edu}

\title{Static Electric Fields as a Model for Hydrogen-Bond-Induced Dissociation of HF and HCl}

\abbreviations{EEF}
\begin{document}

\begin{tocentry}
\includegraphics[scale=0.45]{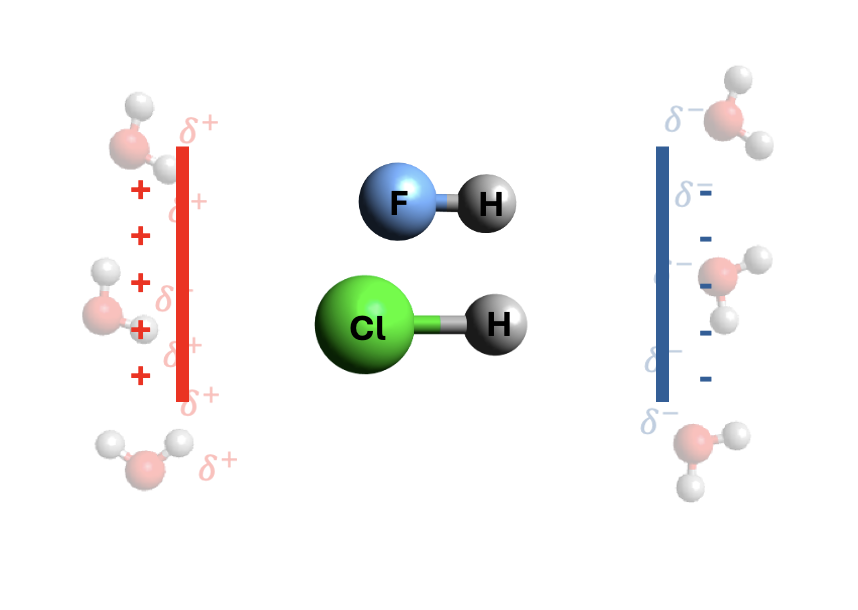}
\end{tocentry}


\begin{abstract}
The influence of static electric fields on the electronic structure and dissociation behavior of the polar diatomics \ce{HF} and \ce{HCl} is investigated using quantum chemical calculations. Ground- and excited-state potential energy surfaces (PESs) are computed as a function of bond distance and external electric field strength to examine field-induced modifications of chemical bonding. The calculations reveal pronounced bond softening and progressive destabilization of both molecules with increasing field intensity. Notably, the ground-state PES of \ce{HCl} becomes entirely dissociative at field strengths of approximately 450 MV/cm, whereas \ce{HF} requires a substantially stronger field of nearly 700 MV/cm to induce dissociation. This difference reflects the greater polarizability and weaker bond localization in \ce{HCl} relative to \ce{HF}, providing a molecular-scale perspective on the contrasting macroscale acid strengths of the two species. Field-dependent dipole moments further demonstrate the stronger electronic response of \ce{HCl} to external perturbations, highlighting how molecular polarizability drives electric-field-induced bond activation. Ultimately, these results map out a detailed picture of field-controlled dissociation in hydrogen halides, supporting the view that local electric fields generated by surrounding hydrogen-bonding networks play a key role in modulating bond activation and condensed-phase acidity.
\end{abstract}


\section{Introduction}
Acid dissociation in aqueous solution plays an important role in chemistry, biology, and materials science.\cite{elementary-processes,elsaesser2002ultrafast} The extent of dissociation depends not only on the intrinsic properties of the solute but also on interactions with the surrounding solvent environment. Hydrogen halides provide simple model systems for investigating the microscopic origins of acid dissociation. In particular, hydrogen fluoride (HF) and hydrogen chloride (HCl) are attractive benchmark molecules because they possess relatively simple electronic structures yet exhibit markedly different acid strengths in aqueous solution. HCl behaves as a strong acid and dissociates almost completely in water, whereas HF is a weak acid that exists in equilibrium between molecular and ionic forms.

Extensive theoretical\cite{lee1996theoretical,smith1999mechanism,conley1999ionic,re1998structures,cabaleiro2002computational,mancini2014effects,forbert2011aggregation,kim2014computational,hollas2015fragmentation,guggemos2015electric,D6CP00530F} and experimental\cite{weimann2002first,farnik2003,masia2007connecting,skvortsov2009hydrated,gutberlet2009aggregation,morrison2010infrared,flynn2010infrared,letzner2013high,zischang2015helium} studies have investigated hydrogen halide--water clusters as microscopic models of acid dissociation. These studies demonstrated that proton transfer depends strongly on the number and arrangement of surrounding water molecules.\cite{rev-hydrated-clusters} For example, isolated HCl-water clusters do not spontaneously produce separated \ce{H3O+} and \ce{Cl-} ions until a sufficient number of water molecules are present.\cite{zwier2009squeezing, mancini2015isolating, vargas2016many, boda2017insights} Previous work estimated that approximately seven, four, three, and three water molecules are required to induce dissociation of HF,\cite{re2001enhanced} HCl,\cite{odde2004dissociation} HBr,\cite{conley1999ionic} and HI,\cite{cabaleiro2002computational} respectively.\cite{rev-hydrated-clusters}

The surrounding solvent molecules generate electric fields that polarize the H-X bond and promote proton transfer.\footnote{The electric field strengths considered in the present work are comparable to those encountered in highly polar condensed-phase environments. For instance, lithium salt solutions have been reported to generate local electric fields of approximately 160~MV/cm.\cite{Hirschfelder1967}} The magnitude and direction of these fields depend on the instantaneous solvent configuration.\cite{boda2017insights,laage2017perspective,cassone2023reactivity} Recent electronic structure and BOMD simulations estimated critical electric fields of approximately 347, 193, 163, and 153~MV/cm for HF, HCl, HBr, and HI, respectively.\cite{pathak2024probing} The critical field decreases systematically from HF to HI, consistent with increasing acid strength across the series. It was further shown that proton transfer in HCl and HBr can occur at substantially lower field strengths in confined environments.\cite{singh2023dissociation}

While these cluster studies focus on explicit molecular counts, an alternative framework treats these local non-covalent environments primarily as electrostatic fields. Boxer and co-workers have established electric fields as a quantitative framework for understanding hydrogen bonding, enzyme catalysis, and solvent effects.\cite{boxer-ef-local, chattopadhyay1995vibrational, park1999vibrational, fried2017electric} With this idea, local non-covalent interactions can be described in terms of the electric field experienced by a molecule rather than solely through structural descriptors.\cite{sen2016internal,boda2020internal}

While hydrated acid clusters have been widely investigated, comparatively little attention has been given to how isolated acid molecules respond directly to external electric fields. Focusing on isolated molecules allows us to characterize the purely electrostatic forces driving dissociation, free from the competing interactions of an explicit solvent network. External electric fields directly perturb molecular electronic structure through the Stark effect, modifying charge distributions, orbital energies, and potential energy surfaces (PESs).\cite{sowlati2013chemical,sowlati2020manipulation} At high field strengths, these perturbations weaken the chemical bond and ultimately drive dissociation.\cite{ramos1997theoretical}

\begin{figure}
\includegraphics[scale=0.7]{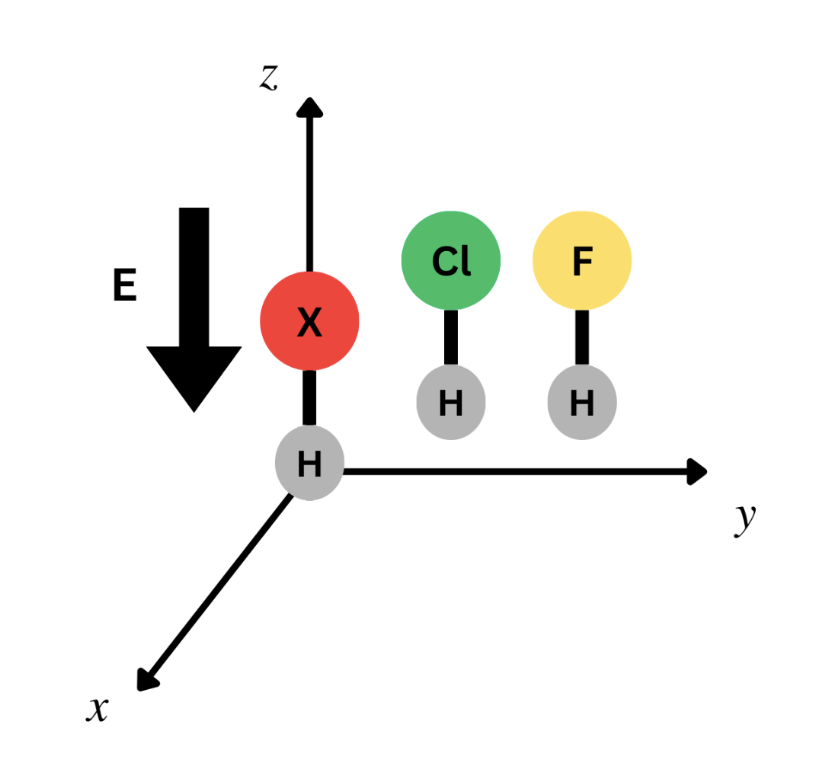}
\caption{Schematic representation of the molecular orientation and direction of the applied external electric field.}
\label{figure:model}
\end{figure}

In this work, we investigate the influence of static electric fields on the electronic structure and dissociation behavior of HF and HCl. The electric field is applied along the molecular axis in the direction opposing the molecular dipole, as illustrated in Figure~\ref{figure:model}. The PESs for the ground and lowest excited electronic states are computed as functions of field strength, together with field-dependent dipole moments and HOMO--LUMO energy gaps. The dependence of the predicted behavior on basis set quality and density functional approximation is examined using the aug-cc-pVXZ ($X=\mathrm{D,T,Q}$) basis sets and the PBE, B3LYP, and BHHLYP functionals. These calculations are complemented by coupled-cluster reference calculations using CCSD and EOM-CCSD methods to assess the robustness of the results across electronic structure levels.

\section{Computational Details}

The equilibrium geometries of HF and HCl were optimized in the absence and presence of external electric fields using density functional theory (DFT).\cite{Hohenberg1964,Kohn1965} The BHHLYP exchange--correlation functional was used for all geometry optimizations.\cite{BHHLYPFaraji} Dipole moments and HOMO--LUMO energy gaps were calculated at the optimized geometries as functions of the applied electric field strength. Ground- and excited-state PESs were computed under field-free and field-applied conditions using DFT and time-dependent density functional theory (TD-DFT). To examine the dependence of the results on the exchange--correlation functional, PES calculations were additionally performed with the B3LYP\cite{becke1993density,lee1988development} and PBE-D4\cite{perdew1996generalized,caldeweyher2019generally} functionals. Ground- and excited-state PESs were also calculated using coupled-cluster theory with single and double excitations (CCSD)\cite{purvisFullCoupledCluster1982,scuseriaClosedshellCoupledCluster1987,leeEfficientClosedshellSingles1988} and equation-of-motion CCSD (EOM-CCSD).\cite{stanton1993equation}

All calculations employed the augmented correlation-consistent polarized valence triple-$\zeta$ basis set, aug-cc-pVTZ.\cite{kendall1992electron,prascher2011gaussian} Basis-set convergence was assessed using the aug-cc-pVXZ family of basis sets ($X=\mathrm{D},\mathrm{T},\mathrm{Q}$). The PESs obtained with aug-cc-pVTZ were in close agreement with those calculated using aug-cc-pVQZ, indicating that aug-cc-pVTZ provides a suitable balance between computational cost and accuracy. All calculations were performed using the ORCA 6.0 software package.\cite{neese2012orca,neese2025software,neese2020orca,neese2022software,neese2023shark} Atomic charges were evaluated using the standard methodology established by Mulliken.\cite{mulliken1955electronic}

\section{Results}

\begin{table}
    \centering
    \begin{tabular}{ccc}
    \hline
    Electric Field (\unit{\mega\volt\per\centi\meter})   & \ce{HF} (\unit{\angstrom}) &  \ce{HCl} (\unit{\angstrom})\\
    \hline
    0   &    0.911  &   1.273 \\
    50  &    0.915  &   1.277 \\
    100 &    0.918  &   1.282 \\
    150 &    0.922  &   1.289 \\
    200 &    0.927  &   1.299 \\
    250 &    0.933  &   1.310 \\
    300 &    0.939  &   1.325 \\
    350 &    0.946  &   1.345 \\
    400 &    0.954  &   1.373 \\
    450 &    0.964  &   1.419 \\
    500 &    0.975  &   -- \\
    550 &    0.989  &   -- \\
    600 &    1.008  &   -- \\
    650 &    1.034  &   -- \\
    700 &    1.086  &   -- \\
    \hline
    \end{tabular}
    \caption{Equilibrium bond lengths of \ce{HF} and \ce{HCl} as a function of static electric field strength.}
    \label{tab:bond-lengths}
\end{table}

\subsection{Equilibrium geometries} 
\label{sec:geometries}

The equilibrium geometries of \ce{HF} and \ce{HCl} were optimized under the influence of static electric fields. The electric field strength was increased in steps of 50 MV/cm. The resulting equilibrium bond lengths are summarized in Table \ref{tab:bond-lengths}. These geometry optimizations resulted in undissociated molecular geometries for electric field strengths up to 700 MV/cm for \ce{HF} and 450 MV/cm for \ce{HCl}. For higher electric field strengths, the field-induced dissociation into \ce{H+} and \ce{X-} species was observed. These results indicate that the critical electric field strength, defined as the field threshold beyond which the isolated molecule dissociates, is 700 MV/cm and 450 MV/cm for \ce{HF} and \ce{HCl}, respectively.

For the electric field strengths below the critical electric field, the equilibrium bond length increases monotonically with increasing electric field strength for both molecules. At low fields (0--200 MV/cm), the bond elongation is approximately linear, with \ce{HCl} exhibiting a larger slope compared to \ce{HF}. At higher fields, deviations from linearity become increasingly pronounced, particularly for \ce{HCl}, indicating enhanced bond softening and approach toward a dissociative regime. At 300 MV/cm, the bond elongation is 0.052 \AA{} for \ce{HCl} compared to 0.028 \AA{} for \ce{HF}.

\subsection{Potential Energy Surfaces}

To understand the effect of external electric fields on bond stability, we calculated the PESs of the ground and low-lying excited electronic states for both \ce{HF} and \ce{HCl}. These electronic structure calculations were performed by changing internuclear distances, $R_{HX}$, under the influence of varying static electric field strengths. The PES for \ce{HF} was evaluated by varying bond lengths, $R_{HF}$ from 0.6 \AA{} to 2.0 \AA{} in steps of 0.05 \AA, while for \ce{HCl} $R_{HCl}$ was varied withing 0.9 \AA{} to 2.0 \AA. 

\begin{figure}
\includegraphics[scale=0.85]{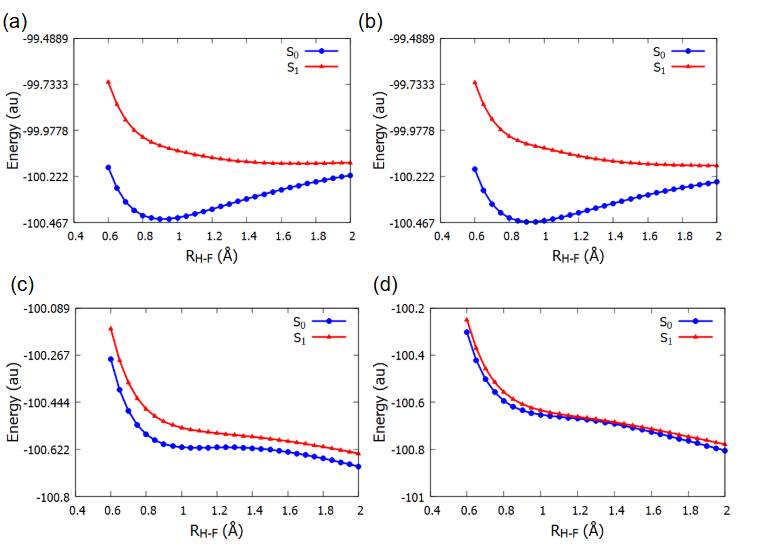}
\caption{Potential energy scans for the ground and lowest excited electronic states of \ce{HF} at electric field strengths of (a) 0 MV/cm, (b) 100 MV/cm, (c) 700 MV/cm, and (d) 800 MV/cm.}
\label{figure:PES-HF}
\end{figure}

\begin{figure}
\includegraphics[scale=0.85]{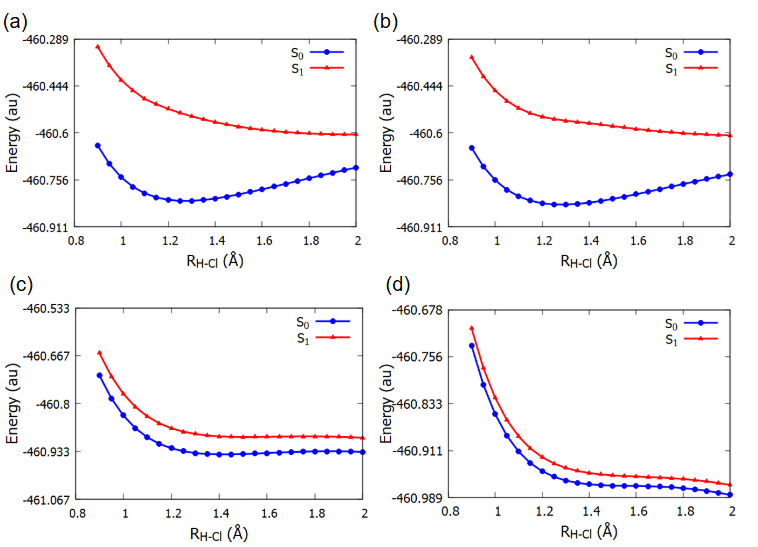}
\caption{Potential energy scans for the ground and lowest excited electronic states of \ce{HCl} at electric field strengths of (a) 0 MV/cm, (b) 100 MV/cm, (c) 450 MV/cm, and (d) 500 MV/cm.}
\label{figure:PES-HCl}
\end{figure}

Figures \ref{figure:PES-HF} and \ref{figure:PES-HCl} show the PESs of the ground and low-lying excited electronic states of \ce{HF} and \ce{HCl} as a function of the internuclear distance under selected static electric field strengths. At zero field (Figures \ref{figure:PES-HF}(a) and \ref{figure:PES-HCl} (a)), both systems exhibit well-defined bound potential wells with clear minima corresponding to equilibrium bond lengths. The excited-state PESs are strictly repulsive, lacking any bound states or stable minima.

As the electric field strength increases, both ground- and excited-state PESs of both \ce{HF} and \ce{HCl} undergo systematic deformation. At a field strength of 100 MV/cm (shown in Figures \ref{figure:PES-HF}(b) and \ref{figure:PES-HCl} (b)), the relative energy difference between the excited state and the ground state as well as the nature of PESs is not affected significantly for \ce{HCl} and \ce{HF}. At 100 MV/cm, the external field is still a weak perturbation compared to the internal electrostatic forces of the molecules. Consequently, the electronic configurations remain mostly unchanged, and both potential surfaces retain their zero-field, gas-phase characteristics.

The transition toward field-induced dissociation is clearly captured at the intermediate field strengths depicted in Figures \ref{figure:PES-HCl}(c) and \ref{figure:PES-HF}(c). At these thresholds—specifically 450 MV/cm for \ce{HCl} and 700 MV/cm for \ce{HF}—the ground-state potential energy surfaces are significantly flattened, retaining only a very shallow local minimum. These shallow wells represent highly destabilized, metastable molecular states on the verge of bond rupture. Concurrently, the ground and lowest-lying excited states begin to exhibit strong interaction in this regime, as evidenced by the narrowing vertical energy gap near the elongated bond regions. This pronounced state-mixing signals the onset of the electronic transition from a covalent description to a highly polarized, field-stabilized ionic state.

Beyond these critical thresholds, the bound molecular states collapse entirely. As shown in Figure \ref{figure:PES-HCl}(d), a field strength of 500 MV/cm renders the ground-state PES for \ce{HCl} strictly repulsive, directly reflecting the absence of a bound equilibrium geometry once the molecule exceeds its critical electric field strength. Furthermore, we observe that at these higher field strengths, the relative energy difference between the excited and ground states decreases significantly. This compression indicates that the applied electric field perturbs the molecular electronic structure in a way that dramatically polarizes the valence electron density, driving the ground state toward a purely ionic, dissociative limit. Similarly, Figure \ref{figure:PES-HF}(d) shows that under the influence of an 800 MV/cm field, the ground-state PES of \ce{HF} becomes entirely repulsive and lacks a bound minimum. For this extreme field strength, the ground and lowest excited states interact strongly and appear to approach degeneracy across a wide range of H–F internuclear distances.

These PES profiles demonstrate that \ce{HCl} has a much stronger response to the external electric field than \ce{HF}, which is evident in both the bound and dissociative regions of the potential curves. These field-dependent changes in the PES directly explain the structural behavior and bond elongation trends described earlier in Section~\nameref{sec:geometries}.

\begin{table}
\caption{Field-dependent partial atomic charges for \ce{HF} evaluated at the equilibrium geometry, R$_{HF}$=0.911\AA.}
\label{table:charges-hf}
\centering
\begin{tabular}{ccc}
\hline
Electric Field (\unit{\mega\volt\per\centi\meter}) & q(H) & q(F) \\
\hline
0   & +0.351 & -0.351 \\
50  & +0.382 & -0.382 \\
100 & +0.412 & -0.412 \\
150 & +0.442 & -0.442 \\
200 & +0.471 & -0.471 \\
250 & +0.500 & -0.500 \\
300 & +0.529 & -0.529 \\
350 & +0.558 & -0.558 \\
400 & +0.586 & -0.586 \\
450 & +0.614 & -0.614 \\
500 & +0.642 & -0.642 \\
550 & +0.668 & -0.668 \\
600 & +0.694 & -0.694 \\
650 & +0.718 & -0.718 \\
700 & +0.740 & -0.740 \\
750 & +0.757 & -0.757 \\
800 & +0.764 & -0.764 \\
850 & +0.764 & -0.764 \\
\hline
\end{tabular}
\end{table}

\begin{table}
\caption{Field-dependent partial atomic charges for \ce{HCl} evaluated at the equilibrium geometry, R$_{HCl}$=1.276 \AA.}
\label{table:charges-hcl}
\centering
\begin{tabular}{ccc}
\hline
Electric Field (\unit{\mega\volt\per\centi\meter}) & q(H) & q(Cl) \\
\hline
0   & +0.171 & -0.171 \\
50  & +0.215 & -0.215 \\
100 & +0.256 & -0.256 \\
150 & +0.295 & -0.295 \\
200 & +0.331 & -0.331 \\
250 & +0.366 & -0.366 \\
300 & +0.399 & -0.399 \\
350 & +0.431 & -0.431 \\
400 & +0.460 & -0.460 \\
450 & +0.488 & -0.488 \\
500 & +0.513 & -0.513 \\
550 & +0.531 & -0.531 \\
\hline
\end{tabular}
\end{table}

\begin{figure}
\includegraphics[scale=0.7]{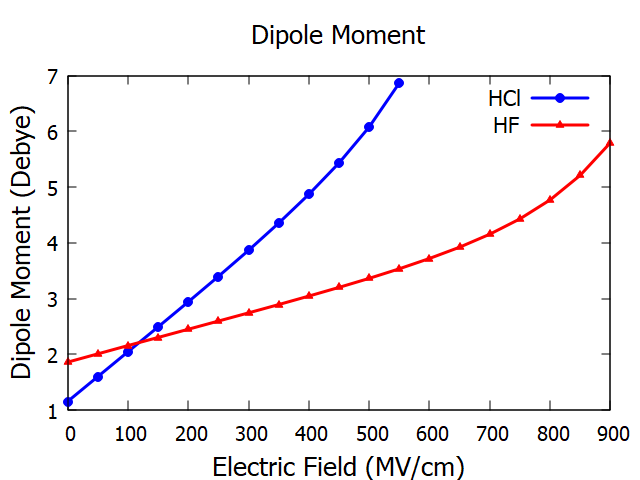}
  \caption{Dipole moments of \ce{HF} and \ce{HCl} molecules evaluated for variable electric field strengths. The geometries of these molecules were fixed to the respective equilibrium bond lengths, R$_{HF}$ = 0.911\AA, R$_{HCl}$ = 1.276 \AA.}
  \label{figure:dm}
\end{figure}

\subsection{Partial Atomic Charges and Dipole Moments}

The field-dependent partial atomic charges for \ce{HF} and \ce{HCl}, calculated within the Mulliken population analysis framework, are presented in Tables~\ref{table:charges-hf} and \ref{table:charges-hcl}, respectively. These calculations were performed at the fixed zero-field equilibrium bond lengths ($R_{\mathrm{HF}} = 0.91$~\AA{} and $R_{\mathrm{HCl}} = 1.276$~\AA{}) to isolate the electronic response from structural geometry relaxation. 

For both molecules, the magnitude of the partial charges increases monotonically with the applied electric field strength. At zero field, \ce{HF} exhibits a higher intrinsic charge separation ($q = \pm 0.351$) compared to \ce{HCl} ($q = \pm 0.171$), reflecting the higher electronegativity of the fluorine atom. As the collinear electric field is applied, electron density is systematically driven from the hydrogen atom toward the halogen center. Consequently, the hydrogen atom becomes increasingly electropositive, while the complementary halogen accumulates an equivalent negative charge, drawing both systems closer toward their ionic limits (\ce{H+} and \ce{X-}).

However, the rate of charge redistribution with increasing field strength ($dq/dE$) differs visibly between the two halides. In the case of \ce{HCl}, the partial charge on hydrogen nearly triples from $+0.171$ at zero field to $+0.513$ at 500~MV/cm. By contrast, the charge on hydrogen in \ce{HF} increases from $+0.351$ to $+0.642$ over the same 500~MV/cm range. This sharper relative increase in \ce{HCl} indicates a more compliant valence electron cloud. The larger, more diffuse valence orbitals of the chlorine atom yield a higher polarizability than the compact, tightly bound valence shell of fluorine, making the electronic structure of \ce{HCl} more susceptible to external electrostatic perturbations.

Figure~\ref{figure:dm} shows the variation of the molecular dipole moments for \ce{HF} and \ce{HCl} as a function of the applied electric field strength. This figure directly corroborates the trends observed in the Mulliken partial charges. Throughout the entire investigated field range, the total dipole moment of \ce{HF} remains higher than that of \ce{HCl}, which is expected given the significantly larger initial charge separation and shorter but highly polar bond of the fluoride system. 

At low field strengths (0 to 200~MV/cm), the dipole moments for both molecules increase approximately linearly, representing a standard perturbative Stark-effect response. As the field strength exceeds this regime, non-linear polarization effects become increasingly evident, particularly for \ce{HCl}. The steeper slope observed for \ce{HCl} across the field range matches its more drastic $dq/dE$ charge separation rate, attributable to its larger electronic polarizability and weaker bond strength. This enhanced dipole response confirms that the valence electron density of \ce{HCl} undergoes much more substantial distortion under external fields than \ce{HF}. This aligns with the earlier observation of breakdown of PES of \ce{HCl} at lower critical field strengths.

\begin{figure}
\includegraphics[scale=0.7]{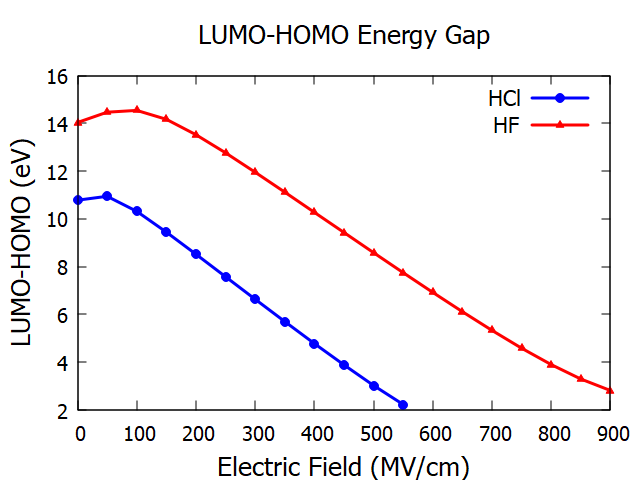}
  \caption{HOMO--LUMO energy gaps for \ce{HF} and \ce{HCl} molecules calculated at variable electric field strengths. The geometries were fixed to the respective equilibrium bond lengths (R$_{HF}$ = 0.911 \AA and R$_{HCl}$ = 1.276 \AA).}
  \label{figure:lumo-homo}
\end{figure}

\subsection{HOMO--LUMO Energy Gaps}

Figure~\ref{figure:lumo-homo} shows the variation of the HOMO--LUMO energy gaps for \ce{HF} and \ce{HCl} as a function of the applied static electric field strength. As with the partial charge and dipole moment analyses, these calculations were performed at the fixed zero-field equilibrium bond lengths of each molecule. 

For both systems, the HOMO--LUMO gap decreases monotonically as the electric field strength increases. At zero field, \ce{HF} possesses a significantly larger fundamental energy gap than \ce{HCl}, which aligns with the shorter, stronger covalent bonding and higher stability of the fluoride molecule. When the external field is applied, it perturbs the molecular orbital energies through the Stark effect, systematically lowering the energy separation between the occupied and virtual spaces.

The rate of this gap narrowing is much more pronounced in \ce{HCl} than in \ce{HF}. As the field increases, the HOMO--LUMO gap of \ce{HCl} drops sharply, which corresponds to the high polarizability of its diffuse valence electron cloud. This rapid compression of the orbital gap directly supports our earlier observations: it explains the heightened non-linear dipole polarization and the lower critical field threshold required to flatten the ground-state potential energy surface of \ce{HCl}. By contrast, the HOMO--LUMO gap of \ce{HF} decreases at a much more gradual rate, confirming that its electronic structure strongly resists external electrostatic distortion up to much higher field thresholds.

\section{Discussion}

The potential energy surface (PES) calculations in the main text were evaluated using the BHHLYP functional. To verify the reliability of these curves, we benchmarked this approach against several electronic structure levels in the Supporting Information (Figures S1–S3), including PBE-D4, B3LYP, and EOM-CCSD methods using the aug-cc-pVTZ basis set. The qualitative features of the PESs remain entirely consistent across all tested density functional approximations and the wave-function-based EOM-CCSD reference. Each method confirms that the ground-state potential well flattens and becomes strictly repulsive at high field strengths, while the excited states remain purely repulsive across all field regimes. 

We also evaluated the basis set dependence of the BHHLYP calculations by comparing the aug-cc-pVTZ results with the smaller aug-cc-pVDZ and larger aug-cc-pVQZ basis sets (Figures S4 and S5). This comparison shows excellent convergence of the potential energy curves, with convergence issues occurred only for certain R$_{HF}$ and R$_{HCl}$ values at field strengths exceeding the critical electric field. Moving from the triple-$\zeta$ to the quadruple-$\zeta$ basis set yields negligible changes in both the predicted equilibrium bond lengths and the critical field thresholds required for dissociation. This demonstrates that the aug-cc-pVTZ basis set balances computational efficiency with a highly accurate description of the electronic structure under intense electrostatic perturbations.

The earlier breakdown of the structure optimization for \ce{HCl} compared to \ce{HF} provides a direct estimate of the relative critical field strengths required for bond dissociation. This trend aligns with classic studies on the influence of external fields on hydrogen-bonded acid-base complexes \cite{ramos1997theoretical}, and highlights how molecular polarizability governs electric-field-driven bond activation. Ramos et al.\cite{ramos1997theoretical} predicted a critical field of 510 MV/cm for \ce{HCl} dissociation, which is in excellent quantitative agreement with our calculated threshold of 450 MV/cm. This consistency underscores the reliability of the BHHLYP/aug-cc-pVTZ framework in capturing intense electrostatic perturbations.

This strong connection between electrostatic forces and chemical reactivity is well-supported across varying scales in recent literature. Boda and Patwari \cite{boda2017insights} previously demonstrated that internal electric fields play a decisive role in driving the acid dissociation of \ce{HCl} and \ce{HBr}. Further, Pathak et al.\ \cite{pathak2024probing} showed that the specific configuration of surrounding solvent networks dictates these local electric fields within explicit water clusters. Notably, Pathak et al.\cite{pathak2024probing} reported critical electric fields 347 MV/cm for \ce{HF} and 193 MV/cm for \ce{HCl} dissociation that are roughly half the magnitude of the thresholds observed in our uniform field calculations. This discrepancy is physically reasonable; their values account for the localized, highly directional electric fields and explicit hydrogen-bonding networks provided by discrete water molecules, whereas our model utilizes a simplified, uniform external field.

Beyond simple aqueous environments, Welborn \cite{vaissier2023understanding} successfully extended this electrostatic paradigm to heterogeneous biological systems, utilizing local electric fields to rationalize dramatic $pK_a$ shifts and reactive states of functional residues inside protein environments. Taken together, these studies confirm that localized reaction fields—whether generated by microhydration shells or complex macromolecular matrices—fundamentally drive bond polarization and alter chemical acidity. Our quantitative trends strongly reinforce this paradigm, validating the use of uniform external electric fields as an effective, physically intuitive proxy for local non-covalent environments.

\section{Conclusions}

In this work, we investigated the effect of static external electric fields on the structural and electronic properties of isolated \ce{HF} and \ce{HCl}. Following the framework established by Boxer and co-workers,\cite{boxer-ef-local} these uniform fields provide a direct way to model the localized electrostatic reaction fields found in hydrogen-bonding and microhydration networks without the complexity of explicit solvent molecules.

Our results show that increasing the external electric field systematically polarizes both diatomic molecules, driving them toward charge-separated, ionic configurations. However, the two systems show a clear difference in their sensitivity to the applied field:
\begin{itemize}
    \item \ce{HCl} responds strongly to the external field, showing large bond elongations, severe flattening of its ground-state potential energy surface, and a sharp narrowing of the HOMO--LUMO gap. This high sensitivity is driven by the large polarizability and weaker bond strength of the chlorine atom, leading to a lower critical dissociation field of 450 MV/cm.
    
    \item \ce{HF} is much more resistant to field-driven electronic distortion. It maintains a well-defined bound potential well up to a much higher critical field threshold of 700 MV/cm, reflecting its compact valence shell and stronger covalent bond.
\end{itemize}

This difference in localized field sensitivity explains the macroscale acid strengths of the two species in solution. The high polarizability of \ce{HCl} allows local solvent environments to easily deform its potential surface and promote proton dissociation, making it a strong acid. Conversely, the electronic rigidity of \ce{HF} resists these external electrostatic forces, explaining why it remains a weak acid despite having a larger intrinsic dipole moment. These findings show that static electric fields can be used as a simple, effective descriptor for understanding non-covalent bond activation and proton-transfer pathways in condensed-phase environments.

\begin{acknowledgement}
Authors thank University of Dayton for the support. This work utilized computational resources provided by the Ohio Supercomputer Center on the Pitzer Cluster under allocation PNS0499. The authors gratefully acknowledge this support.
\end{acknowledgement}


\begin{suppinfo}
Electronic supplementary information files are available free of charge at the publisher's website. Supporting\_Information.docx includes the comparison of the results obtained for Potential Energy Surfaces evaluated with different density-functionals, EOM-CCSD method and basis sets. 
\end{suppinfo}

\bibliography{bibliography}

\end{document}